\documentclass[sigconf,nonacm]{acmart}

\usepackage{listings}
\definecolor{tvlbg}{HTML}{F6F6F6}
\definecolor{tvlrule}{HTML}{B8B8B8}
\definecolor{tvlblock}{HTML}{1D3557}
\definecolor{tvlfield}{HTML}{2B6E4E}
\definecolor{tvlstring}{HTML}{9B2226}

\lstdefinelanguage{tvlmini}{
  alsoletter={_,-},
  keywords={tvl,environment,evaluation_set,tvars,constraints,structural,objectives,promotion_policy},
  morekeywords=[2]{module,models,budget_usd,dataset,seed,name,type,domain,when,then,direction,measure,test,dominance,alpha,min_effect,tie_breakers,chance_constraints,threshold,confidence},
  morekeywords=[3]{maximize,minimize,epsilon_pareto,enum,bool,int,true,false},
  sensitive=true,
  morestring=[b]",
}

\lstdefinestyle{tvlstyle}{
  language=tvlmini,
  basicstyle=\ttfamily\scriptsize,
  keywordstyle=\color{tvlblock}\bfseries,
  keywordstyle=[2]\color{tvlfield},
  keywordstyle=[3]\color{blue!60!black},
  stringstyle=\color{tvlstring},
  backgroundcolor=\color{tvlbg},
  frame=l,
  framerule=1pt,
  rulecolor=\color{tvlrule},
  xleftmargin=8pt,
  breaklines=true,
  columns=fullflexible,
  keepspaces=true,
  showstringspaces=false,
  numbers=none,
  aboveskip=4pt,
  belowskip=2pt,
}

\title{Statistical Software Engineering with Tuned Variables}

\author{Nimrod Busany}
\affiliation{\institution{Traigent}\city{Tel Aviv}\country{Israel}}

\begin{document}

\begin{abstract}
The maintained artifact in an AI-enabled system is not code plus settings, but a versioned governed program space: domains, structural constraints, eligibility, evaluation assets, and a statistical release gate. AI-enabled systems operate under changing world conditions: provider models and APIs change, input distributions drift, evaluation sets age, and objectives such as quality, cost, latency, and safety are renegotiated over time~\cite{sculley2015debt,gama2014drift}. In practice, teams often respond through ad hoc changes to model choice, retrieval policy, prompt structure, and operational thresholds. Fixed-assignment reasoning is therefore insufficient: a chosen assignment is valid only relative to an environment, evaluation set, and policy state. We argue that such choices should be treated as \emph{tuned variables}: program variables maintained under governance as environments and evaluation sets evolve. Building on SE4AI work~\cite{amershi2019se4ml} and our prior work on governed tuning~\cite{busany2026tvo}, this paper positions the governed space as the software-engineering object. Here, \emph{statistical} means that promotion relies on sampled evaluation sets, estimated evidence, effect-size margins, and confidence/risk thresholds.
\end{abstract}

\maketitle

\section{Introduction}
Classical software engineering has powerful abstractions for code, tests, and releases, but AI-enabled systems expand and destabilize the governable surface. Available models and APIs, input distributions, evaluation sets, and product objectives all change over time. As a result, choices over model, prompt form, retrieval depth, tool usage, and thresholds must be repeatedly validated with respect to accuracy, cost, latency, robustness, and safety.

LLM applications show that alternative assignments can differ materially in cost and quality~\cite{busany2025eco}. SE4AI work characterizes ML systems through technical-debt, process, and survey lenses~\cite{sculley2015debt,amershi2019se4ml,martinez2022survey}; Lo's roadmap frames trustworthy AI4SE as a field-level transformation~\cite{lo2023trustworthy}. SBSE frames software problems as search~\cite{harman2012sbse}, and DSPy/TextGrad optimize LLM pipelines against metrics~\cite{khattab2024dspy,yuksekgonul2024textgrad}. Statistical Software Engineering (SSE) instead frames maintenance of a governed program space under structural constraints, eligibility, evaluation sets, and release policy.

The distinction is artifact-level. In contrast to~\cite{busany2026tvo}, this paper (i) gives an operational test for tuned variables; (ii) treats eligibility as a property of whole program variants rather than isolated variables; (iii) frames promotion as a statistical multi-objective decision with safety constraints; and (iv) names the governed space itself, not the lifecycle acting on it, as the versioned SE artifact. TVL is only the declaration notation used below.

DSPy, TextGrad, and AutoML~\cite{khattab2024dspy,yuksekgonul2024textgrad,hutter2019automl} focus on optimizing assignments; SSE treats the governed space over which such optimization is admissible. SPLs and configurable systems~\cite{clements2001spl,benavides2010featuremodels,xu2015knobs} define valid variant spaces, but do not adjudicate admission; SSE's $G$ does. A/B testing~\cite{kohavi2020online} adjudicates live traffic alternatives after deployment; SSE gates admission into that population beforehand. MLOps registries center on model assets~\cite{kreuzberger2023mlops}; SSE's governed object spans prompts, retrieval, tools, and policy. NRC 1996 enunciated statistical software engineering for classical software~\cite{nrc1996sse}; here, tuned-variable assignments denote candidate program variants, not settings files.

\section{Running Example}
\label{sec:example}
Consider a support-assistant pipeline. Engineers may choose the provider model, retrieval depth, prompt template, whether conversation history is used, and how many prior messages ($k$) are included. Some assignments are structurally impossible: if history is disabled, then $k$ must be zero. Others are inadmissible in the current environment because a model is unavailable or a cost cap is violated. The goal is therefore not to find one timeless setting, but to maintain a governed space of admissible program variants and decide, with evidence, when a candidate may replace the incumbent. \textsc{TVL}~\cite{tvl2026} is used here only as a compact declaration of that space:
\begin{lstlisting}[style=tvlstyle]
tvl: { module: support.assistant }
environment: { models: [gpt-5.4-mini, gpt-5.4, claude-opus-4-7], budget_usd: 0.02 }
evaluation_set: { dataset: support_tickets_v3, seed: 13 }
tvars:
  - { name: model, type: enum[str], domain: environment.models }
  - { name: retrieval_depth, type: int, domain: [4, 8, 12] }
  - { name: prompt_template, type: enum[str], domain: [brief, guided] }
  - { name: history, type: bool, domain: [true, false] }
  - { name: k, type: int, domain: [0, 2, 4, 6, 8] }
constraints: { structural: [{ when: history = false, then: k = 0 }] }
objectives:
  - { name: quality, measure: answer_accuracy, direction: maximize }
  - { name: cost, direction: minimize }
  - { name: latency, direction: minimize }
promotion_policy:
  { dominance: epsilon_pareto, alpha: 0.05,
    min_effect: { quality: 0.01 },
    tie_breakers: [cost, latency],
    chance_constraints: [{ name: bias_rate, test: fairness_test, threshold: 0.05, confidence: 0.95 },
                         { name: hallucination_rate, test: hallucination_check, threshold: 0.03, confidence: 0.95 }] }
\end{lstlisting}

The next sections use this example to define the notation.

\section{Tuned Variables, Eligibility, and Pareto Assignment}
\noindent\textbf{Definition 1 (Tuned variable).} A program variable $t$ is tuned when it affects externally evaluated behavior, its admissible domain may change with the environment, changes require statistical re-validation, not code review alone, and its assignments are versioned as part of the governed program space. At state $\tau$ (a versioned snapshot of the space), let $\mathcal{V}_\tau=\{t_1,\ldots,t_n\}$ and $\Theta_\tau=D_\tau(t_1)\times\cdots\times D_\tau(t_n)$ denote the assignment space; below, subscripts are dropped when $\tau$ is fixed. In Sec.~\ref{sec:example}, \texttt{model}, \texttt{retrieval\_depth}, \texttt{prompt\_template}, \texttt{history}, and \texttt{k} instantiate $\mathcal{V}$, while the listed domains instantiate $D$. Structural constraints define a predicate $C_s(c)$ over assignments, excluding incompatible permutations before execution; e.g., the running example uses \texttt{history=false} $\Rightarrow$ \texttt{k=0}. For an environment snapshot $e$, eligibility is an assignment-level predicate:
\[
\mathcal{E}_e(c).
\]
Per-variable eligible domains are a common factorization, but realistic environments often constrain combinations of choices; in Sec.~\ref{sec:example}, available models and the cost cap instantiate $\mathcal{E}_e$. The governed feasible set is therefore
\[
\mathcal{F}_e=\{c\in\Theta \mid C_s(c)\wedge \mathcal{E}_e(c)\}.
\]
Following prior work on statistical analysis of LLM evaluations~\cite{miller2024errorbars}, for an evaluation set $S$ the evaluation function of assignment $c\in\mathcal{F}_e$ is
\[
\operatorname{ev}(c,S)=\langle J(c,S),R(c,S)\rangle,
\]
where $J(c,S)=(\phi_1(c,S),\ldots,\phi_p(c,S))$ denotes objective scores and $R(c,S)=(\psi_1(c,S),\ldots,\psi_q(c,S))$ denotes safety measures. In Sec.~\ref{sec:example}, answer accuracy, cost, and latency instantiate $J$, while bias and hallucination rates instantiate $R$~\cite{liang2022helm}.
Each $c\in\Theta$ denotes a program variant. This separates structural validity, environmental admissibility, and promotion. This emphasis on evaluation sets is also consistent with recent evidence that software-engineering evaluation assets are uneven in coverage, realism, and contamination risk~\cite{hu2025benchmarks}; ``repeatable'' here means maintained and refreshable rather than static.

In many AI-enabled systems, the goal is not a single scalar optimum, but a Pareto-efficient assignment over quality, cost, and latency, subject to safety measures. Pareto analysis identifies admissible trade-offs, not the release rule itself. Let $G$ denote the promotion gate: it maps incumbent/candidate evidence and a policy $\pi$ to $\mathsf{pass}$, $\mathsf{defer}$, or $\mathsf{fail}$; the example's \texttt{promotion\_policy} instantiates $\pi$. If $c_0$ is incumbent, $c_1$ candidate, $E_i=\langle J(c_i,S),R(c_i,S)\rangle$, and $\pi=(\varepsilon,\alpha,\delta,\prec,\rho)$ encodes dominance tolerance, risk level, min-effect margins, tie breakers, and chance constraints, then $G(E_0,E_1,\pi)=\mathsf{pass}$ iff the candidate $\varepsilon$-Pareto-dominates the incumbent on $J$ with margins $\delta$ at confidence $1-\alpha$ and all constraints $\rho$ on $R$ hold; $G=\mathsf{defer}$ if evidence is inconclusive, and $\mathsf{fail}$ otherwise. Promotion is therefore a governed transition. \emph{Observation.} Tightening $C_s$ or $\mathcal{E}_e$ while holding $S$ and $G$ fixed cannot expand the promotable set.

\section{Lifecycle, Triggers, and Promotion}
Recent work describes a CI/CD-oriented lifecycle in which candidate assignments are proposed within guardrails, evaluated on an evaluation set, checked by policy, and adopted with rollback support~\cite{busany2026tvo}. The position advanced here is that such a lifecycle is justified whenever the maintained governance artifact changes. A compact way to describe a governance state is
\[
\Gamma_{\tau}=\langle D_\tau,C_{s,\tau},\mathcal{E}_{e_\tau},S_\tau,G_\tau\rangle.
\]
Any change to a component of $\Gamma_\tau$ creates a new governance state. Re-evaluation may be triggered when $D_\tau$ changes, $C_{s,\tau}$ changes, $\mathcal{E}_{e_\tau}$ expands or narrows, $S_\tau$ is refreshed, $G_\tau$ changes objectives or risk tolerances, or a scheduled review epoch $r\in\mathcal{T}_{\mathrm{review}}$ is reached. Monitoring-observed drift is a further trigger when no explicit artifact edit has occurred.

Re-evaluation is component-dependent: if $D_\tau$ expands, the incumbent remains valid in the existing subspace while candidates may exploit enlarged domains; if $\mathcal{E}_{e_\tau}$ narrows or $C_{s,\tau}$ invalidates the incumbent, a compliant fallback is required; if $S_\tau$ changes, the incumbent is re-evaluated before candidate comparison; if $G_\tau$ changes, existing evidence is re-adjudicated under the new policy before new search begins.
For the support assistant, promotion is one component of the lifecycle. Suppose the incumbent scores 0.840 on 500 locked promotion tickets and $\delta_{\mathrm{quality}}=0.010$:
\begin{center}
\scriptsize
\setlength{\tabcolsep}{2.8pt}
\begin{tabular}{@{}llll@{}}
$c$ & quality evidence & safety/cost evidence & $G$\\
\hline
$c_1$ & $\Delta=.020$, CI $[.012,.028]$ & bias CI $[.038,.062]>.05$ & $\mathsf{fail}$\\
$c_2$ & $\Delta=.005$, CI $[-.003,.013]$ & CI crosses 0; below $\delta$ & $\mathsf{defer}$\\
$c_3$ & $\Delta=.018$, CI $[.011,.025]$ & safety bounds hold; cost/latency ok & $\mathsf{pass}$\\
\end{tabular}
\end{center}
The rows illustrate that promotion is not ``pick the highest accuracy.'' Candidate $c_1$ appears attractive on quality: its estimated gain is .020 and the interval is above the .010 effect-size margin. Yet its bias-rate interval crosses the .05 safety threshold, so the gate returns $\mathsf{fail}$. Candidate $c_2$ is safe, but its quality interval does not clear $\delta$ at confidence $1-\alpha$; the right outcome is $\mathsf{defer}$ rather than a noisy promotion. Candidate $c_3$ clears the quality margin, keeps the safety bounds within policy, and remains acceptable on cost and latency, so $G$ returns $\mathsf{pass}$. The table therefore shows why promotion is a governed transition over evidence, not a scalar optimization result.

\section{Scope and Research Agenda}
The statistical commitment is modest but essential: evaluation sets are sampled from a broader task population~\cite{miller2024errorbars}, $\operatorname{ev}(c,S)$ is estimated with finite uncertainty, and weak evidence should preserve the incumbent. Making $G$ concrete under noisy AI evaluation is beyond this position paper. The governed-space view instead exposes release-gate requirements: paired uncertainty over objective deltas, bounded evidence for low-rate safety measures, multiplicity and selection effects across candidates, and incumbent re-evaluation when governance-state changes affect evidence, policy, or admissibility.

Governed program spaces are large, conditional, and noisy; practical search is budgeted and approximate. This exposes three SE research problems: gate semantics for pass/fail/defer, feasibility estimation under $C_{s,\tau}\wedge\mathcal{E}_{e_\tau}$ without exhaustive evaluation, and evaluation-asset governance across $\tau\!\to\!\tau+1$ for refresh, contamination, drift, and incumbent re-evaluation. SSE names this maintenance view.

\end{document}